%Paper: hep-th/9304130
%From: zaslow@string.harvard.edu (Eric Zaslow)
%Date: Mon, 26 Apr 93 22:35:14 -0400

% This paper is in plain TeX and uses harvmac macros.
% Two unessential figures not included but are sketched below.
%
%
%             o - o - o
%            /         \
%           o           o       Figure One
%            \         /
%             o - o - o
%
%
%         o               o
%          \             /
%           o - o - o - o       Figure Two
%          /             \
%         o               o
%

\input harvmac.tex
\def\half{{1\over 2}}
\def\sub{\scriptscriptstyle}
\def\app#1#2{\global\meqno=1\global\subsecno=0\xdef\secsym{\hbox{#1.}}
\bigbreak\bigskip\noindent{\bf Appendix.}\message{(#1. #2)}
\writetoca{Appendix {#1.} {#2}}\par\nobreak\medskip\nobreak}
%
%       \eqn\label{a+b=c}	gives displayed equation, numbered
%				consecutively within sections.
%     \eqnn and \eqna define labels in advance (of eqalign?)
%
\def\eqnn#1{\xdef #1{(\secsym\the\meqno)}\writedef{#1\leftbracket#1}%
\global\advance\meqno by1\wrlabeL#1}
\def\eqna#1{\xdef #1##1{\hbox{$(\secsym\the\meqno##1)$}}
\writedef{#1\numbersign1\leftbracket#1{\numbersign1}}%
\global\advance\meqno by1\wrlabeL{#1$\{\}$}}
\def\eqn#1#2{\xdef #1{(\secsym\the\meqno)}\writedef{#1\leftbracket#1}%
\global\advance\meqno by1$$#2\eqno#1\eqlabeL#1$$}
\Title{HUTP-93/A010}{\vbox{\centerline{Dynkin Diagrams of}
\vskip2pt\centerline{${CP}^1$ Orbifolds}}}

\centerline{Eric Zaslow\footnote{$^*$}
{Supported in part by Fannie and John Hertz Foundation.}}

\bigskip\centerline{Lyman Laboratory of Physics}
\centerline{Harvard University}\centerline{Cambridge, MA 02138}

\vskip .3in

We investigate $N=2$ supersymmetric sigma model orbifolds of the sphere
in the large radius limit.  These correspond to $N=2$ superconformal
field theories.  Using the equations of topological-anti-topological
fusion for the topological orbifold, we compute the generalized Dynkin
diagrams of these theories - i.e., the soliton spectrum - which was used in the
classification of massive superconformal theories.  They correspond to
the extended Dynkin diagrams associated to finite subgroups of $SO(3).$

\Date{4/93}

\newsec{Introduction and Summary}
\nref\rCVNEW{S. Cecotti and C. Vafa, ``On Classification of $N=2$
Supersymmetric Theories,'' Harvard and SISSA preprints HUTP-92/A064 and
SISSA-203/92/EP.}
\nref\rCHSW{P. Candelas, G. Horowitz, A. Strominger, and E. Witten,
Nucl Phys {\bf B258} (1985) 46.}
\nref\rDHVW{L. Dixon, J. Harvey, C. Vafa, and E. Witten, Nucl. Phys. {\bf
B261} (1985) 678, and Nucl. Phys. {\bf B274} (1986) 285;
L. Dixon, D. Friedan, E. Martinec, and S. Shenker, Nucl.
Phys. {\bf B282} (1987) 13;
and D. Freed and C. Vafa, Comm. Math Phys. {\bf 110} (1987) 349.}
\nref\rZAZ{E. Zaslow, ``Topological Orbifold Models and Quantum Cohomology
Rings,'' Harvard preprint HUTP-92/A065, to appear in Communications in
Mathematical Physics.}
\nref\rCVMO{S. Cecotti and C. Vafa, ``Massive Orbifolds," SISSA and Harvard
preprint SISSA 44/92/EP, HUTP-92/A013.}
\nref\rCVTAF{S. Cecotti and C. Vafa, Nucl. Phys {\bf B367}
(1991) 3543.}
\nref\rJR{R. Jackiw and C. Rebbi, Phys. Rev. {\bf D13} 3398-3409 (1976);
J. Goldstone and F. Wilczek, Phys. Rev. Lett.
{\bf 47} 986-989 (1981).}
\nref\rWW{X.-G. Wen and E. Witten, Nucl. Phys. {\bf B261} (1985) 651.}
\nref\rHV{S. Hamidi and C. Vafa, Nucl. Phys. {\bf B279} (1989) 465.}
\nref\rM{G. Mackey, Acta Math. {\bf 99} (1958) 265.}
\nref\rCVISING{S. Cecotti and C. Vafa, ``Ising Model and $N=2$ Supersymmetric
Theories," Harvard and SISSA preprints HUTP-92/A044, SISSA-167/92/EP.}
\nref\rSLOD{J. McKay, {\sl Cartan Matrices, Finite Groups of Quaternions, and
Kleinian Singularities,} Proc. Am. Math. Soc. (1981) 153; P. Slodowy, {\sl
Simple Singularities and Simple Algebraic Groups,} Lecture Notes in Mathematics
815, Springer-Verlag, New York 1980.}
\nref\rPIII{A. R. Its and V. Yu. Novokshenov, {\sl The Isomonodromic
Deformation Method in the Theory of Painlev\'e Equations,} Lecture Notes
in Mathematics 1191, Springer-Verlag, New York 1986; B.M. McCoy, C.A.
Tracy, and T.T. Wu, J. Math Phys. {\bf 18} (1977) 1058.}
\nfig\fan{The double cover of
$Z_N$ is $Z_{2N},$ from which we determine the associated
affine Lie algebra.  We find $Z_N \leftrightarrow \hat{A}_{2N-1}.$
The case $N = 4$ is shown.}
\nfig\fdi{The extended
Dynkin diagram of the affine Lie algebra
$\hat{D}_7.$ This corresponds to the dihedral group
$D_5.$}

The recent classification of $N=2$ superconformal field theories with
massive deformations \rCVNEW\ provides a major step towards zeroing in on
phenomenologically desirable string vacua \rCHSW.  Orbifolds are known to
be consistent string vacua as well \rDHVW.  The superconformal orbifold
theories considered here are constructed by taking the large radius (Ricci
flat) limit
of massive theories - supersymmetric sigma model orbifolds of the sphere (${\bf
CP}^1$).  Topological orbifolds of sigma
models were described in \rZAZ.  The theories we consider are asymptotically
free.  As such, they are somewhat less interesting than the Calabi-Yau spaces.
However, their classification reveals some structure.  We will analyze the
topological sectors of these theories to compute the soliton spectrum.  This is
then used to classify the theory.  Specifically, the numbers of solitons
connecting two ground states can be organized in a matrix.  The classification
program puts Diophantine constraints on the allowable integer matrices.  These
constraints are satisfied, for example, by the Cartan matrices of Lie groups
\rCVNEW.  We can thus ask the question:  which matrices are associated to the
${\bf CP}^1$ orbifolds?  We find for the dihedral orbifolds that these matrices
correspond to associated affine Lie groups, and expect the same of the
exceptional cases (for the cyclic case, see \rCVNEW \rCVMO).

We first briefly review topological orbifolds on sigma models (section two) and
the classification of ``massive'' $N=2$ superconformal theories (section
three).  We then discuss dihedral orbifolds of ${\bf CP}^1$ (section four) and
solve the equations of
topological-anti-topological fusion ($tt^*$) \rCVTAF\ to compute the
soliton spectrum for orbifolds of ${\bf CP}^1$ by discrete subgroups of
$SO(3)$, and show that the generalized Dynkin diagrams are those of the
associated affine Lie algebras (section five).  As an example, the $D_5$
orbifold is computed in the appendix.

\newsec{Topological Orbifolds}

The topological field theories associated to orbifolds of sigma models
were discussed in \rZAZ.  In the usual topological sigma model
associated to a Kahler manifold, $K$, of complex dimension $d$, the
observables are the cohomology classes.  The Hodge grading corresponds
to chiral fermion numbers $(p,q)\leftrightarrow(f_L,d-f_R).$
In the orbifold model, in which we have an action of a group, $G,$ by
holomorphic
isometries of $K,$ we
project to group invariant cohomology classes.
In addition, we have twisted operators, which create twisted states from
the NS vacuum.  These observables are described by the cohomology classes
of the manifolds fixed under the group elements.  So, for example, the
$g-$twisted observables coincide with $H^*(M_g),$ where $M_g = \lbrace m
\in K \vert gm = m \rbrace.$  If the group is nonabelian, $g$ is
understood to represent a conjugacy class; all $M_g$ are homeomorphic for
a given conjugacy class, so any representative suffices (the actual
observables are appropriate invariant combinations of operators
associated to differential forms for the various
$M_g, g \in \lbrace g \rbrace ).$
The fermions of the theory have tangent space indices, so at fixed
points, where there is a non-trivial action of the group on the normal
bundle to the manifold, they obey twisted boundary conditions.  The
fermionic vacuum, in such a case, undergoes a chiral fermion number shift
\rJR \rWW.  If we consider the observables as
differential forms on the orbifold, the grading of the twisted forms gets
shifted.  This is expressed in the compact formula
\eqn\twistcohom{H^{p,q}(K/G) \equiv \bigoplus_{\lbrace g \rbrace}
H^{p-F_g,q-F_g}_{C(g)}(M_g),}
where $F_g$ represents the fermion number shift, and is given by the sum
of the phases of the eigenvalues of the $g$ action on the normal bundle
of $M_g.$  $C(g)$ is the centralizer of $g,$ i.e. $C(g) = \{h\in G \vert hg =
gh\}.$  Here $p$ and $q$ can be fractional, but $p-F_g$ and $q-F_g$
are integers.

The ring of observables - which coincides with the chiral ring of the
nontopological
theory - can be calculated from the topological three point functions,
which are calculated from an appropriate moduli space.  In the usual
sigma model, this space is just the set of holomorphic maps from the
world sheet to target space.  In the orbifold case, the notion of
holomorphic maps is ill-defined, since the orbifold is not a manifold.
Instead, we take equivariant holomorphic maps with respect to an
appropriate branched cover of the sphere (see
\rHV).\footnote{$^\dagger$}{Throughout this
paper, we take the philosophy of defining the theory through
factorization; thus we only consider genus zero correlation functions.
This avoids problems of compactifying moduli space for higher genus
maps.}  That is, we choose a Riemann surface, $\widetilde{\Sigma},$ which is a
branched cover by
$G$ over the world sheet $\Sigma.$  The insertion points of operators
correspond to the branch points, and the branching elements correspond to
the twisted sector of the operator.
Then, equivariant maps obey $g\phi(x) = \phi(gx)$ for all
group elements $g.$  They uniquely define maps from the sphere to the
orbifold, which are analytic on nonsingular regions (see below).
$$
\def\mapdown#1{\Big\downarrow\rlap{$\vcenter{\hbox{$#1$}}$}}
\def\downmap#1{\llap{$\vcenter{\hbox{$#1$}}$}\Big\downarrow}
\matrix{
&\widetilde{\Sigma}&\matrix{holomorphic \cr
\overrightarrow{\,\,\,\,\,equivariant\,\,\,\,\,}}&K\cr
&\downmap{G}&&\mapdown{G}\cr
&\Sigma&\matrix{\,\cr\overrightarrow{``holomorphic"\,\,}}&K/G&\cr}$$
The correlation
functions are a sum over contributions from the different components of
instanton moduli space (holomorphic, equivariant maps).

\newsec{Classification of $N=2$ Superconformal Theories}

We briefly review the classification by Cecotti and Vafa
of $N=2$ superconformal theories with
massive deformations \rCVNEW, and its relation to the $tt^*$ equations \rCVTAF.

Any $N=2$ theory yields a topological theory from the $Q-$closed modulo
$Q-$exact observables.  Likewise, the $CPT$ conjugate fields create an
``anti-topological'' theory.  The states of the toplogical theory are
denoted $\vert a \rangle,$ and correspond to the observables $\phi_a.$
The anti-topological states are related by a change of basis, effected
by the {\sl real structure} matrix: $\vert \overline{a}\rangle =
M_{\overline{a}}^b \vert b \rangle.$
The quantum field theory defines a metric on the Hilbert space,
${\cal H}$, which descends to a metric on the topological theory,
since $Q-$exact terms are zero in correlators:
$$g_{a\overline{b}} = \langle \overline{b} \vert a \rangle.$$
There is also a topological metric defined by intersections in an
appropriate moduli space.  These structures are defined for any $N=2$
theory, and become geometrical structures on the space of theories.
We can coordinatize this space by coupling constants $\{ t_i \}.$
Choosing an action $S_0,$ we write $$S(t) = S_0 + \left[{\int d^2\theta \, t_i
\phi_i + c.c.}\right].$$
At each $t,$ we have a chiral ring, isomorphic to the Ramond ground states
of the theory.  We thus have a vector bundle - the bundle of ground states -
with the metric given above
(now $t-$dependent).  A ground state, characterized by its $U(1)$ charge, is
then a section of this bundle;
its wave function, then, is $t-$dependent, and we can thus consider
the connection defined by
$$(A_i)_{a\overline{b}} = \langle \overline{b} \vert \del_i \vert a
\rangle.$$  Then $D_i = \del_i - A_i.$
In fact we can consider a family of connections indexed by a ``spectral
parameter,'' $x:$
$$\nabla_i = D_i - xC_i,$$
$$\overline{\nabla}_{\overline{i}} = \overline{D}_{\overline{i}} -
x^{-1}\overline{C}_{\overline{i}},$$
where $C_i$ represents the action of
$\phi_i;$ that is, $\phi_i\phi_j = (C_i)_j^k\phi_k$
($\overline{C}_{\overline{i}}  = gC_{i}^{\dagger}g^{-1}).$
The $tt^*$ equations, conditions on the metric and the $C_i,$ are then
summarized by the statement that
$\nabla$ and $\overline{\nabla}$ are flat for all $x.$

The solutions to the $tt^*$ equations encode the number of
solitons which saturate the Bogolmonyi bound and connect the ground
states.  In the Landau-Ginzburg case, the soliton numbers have a
topological description in terms of intersection numbers of vanishing
cycles over families of varieties.  The vacua correspond to points
satisfying $dW = 0,$ and solitons connecting them travel along straight
lines in the $W$ plane.  The inverse image of the values of $W$ near a
critical point form spheres in ${\bf C}^n.$  When these spheres
intersect, one can build a soliton path between the vacua.  Because of
this interpretation, the soliton numbers must behave as the intersection
numbers under $t-$dependent perturbations of the superpotential, in
particular when the
vacua become colinear via the perturbation.  In \rCVNEW, the authors
developed the analog of this interpretation for a general $N=2$ theory.

As we saw above, the $tt^*$ equations can be formulated as flatness conditions
on a family
of connections.  The equations have the built-in requirement that the
hermitian metric is independent of the overall phase of the generalized
superpotential.  By generalized superpotential, we mean the values $w_a$
which can be assigned to the different vacua such that the
Bogolmonyi soliton masses (the central terms of the $N=2$ algebra) are
given by the differences of the $w_a$.  These are the canonical
coordinates.  That this independence should hold
follows from the freedom to redefine the phases of the fermions.  The
equations are given in terms of the connections
\eqn\conns{\eqalign{\nabla_i &= \del_i + (g\del_i g^{-1}) - xC_i, \cr
\overline{\nabla}_{\overline{i}} &= \overline{\del}_{\overline{i}}
- x^{-1}\overline{C}_{\overline{i}},}}
written here in the $A_{\overline{i}} = 0$ gauge.  We
consider the set of equations
\eqn\lax{\nabla_i \Psi(x,w_a) =
\overline{\nabla}_{\overline{i}}\Psi(x,w_a) = 0.}
In order to solve these equations simultaneously, we must require that
$\nabla$ and $\overline{\nabla}$ commute, i.e. they are {\sl flat};
this consistency condition is $tt^*$.  In general, there will be $n$
solutions to \lax, so we take $\Psi$ to be an $n\times n$ matrix whose
columns are solutions.  The
equations are singular at $x=0,\infty,$ which means the columns of $\Psi$
will mix under monodromy $x \rightarrow \e{2\pi i}x:$  $\Psi \rightarrow
H\cdot\Psi.$  If we consider $\beta \rightarrow
0$ with $x$ small, i.e. the conformal limit,
then these equations indicate that the phases eigenvalues of
the monodromy around zero are precisely the Ramond charges.  Since
these charges must be real, the eigenvalues $\lambda_i = \e{2\pi iq_i}$
of the monodromy must satisfy $\vert \lambda_i \vert = 1.$  Because the
equations of $tt^*$ are flatness equations, they describe isomonodromic
deformations.  That is, the monodromy is a constant.  Indeed it is
calculable in the $\beta \rightarrow \infty$ limit, where the monodromy
$H$ of $\Psi$ is expressable in terms of the soliton numbers $A_{ij}.$
The relation is
\eqn\monod{\eqalign{H &= S(S^{-1})^t, \cr S &= 1 - A.}}
Statements about the charges (e.g. $CPT,$ unique highest/lowest charge vacua)
are then conditions on the possible matrices
$A.$  This is detailed in section six of \rCVNEW.  To us, the important result
is that simply laced Lie groups lead to solutions to the Diophantine equations
of classification.

The simply laced Lie groups are related to possible solutions for $A$ as
follows.  Suppose the matrix $B = S + S^t$ is positive definite.  Then
$HBH^t = SS_{-t}(S + S^t)S^{-1}S^t = B,$ which means that $H$ is in the
orthogonal group to the quadratic form, $B,$ which tells us that $H$ is
simple and $\vert \lambda_i \vert = 1.$  The simply laced Lie groups
correspond to positive definite integral matrices by constructing a
Dynkin diagram from the matrix.
$B$ defines an inner product
on ${\bf R}^n,$ and if we take $A$ to be upper triangular, with
$A_{ij} = -B_{ij}/2, i < j,$ then $H = (1-A)(1-A)^{-t}$ satisfies the
Diophantine
constraints.  These matrices correspond to the $N=2$ $A-D-E$ minimal models.
Weyl reflections of the lattice vectors produce different, though
equivalent solutions to the Diophantine equations.  These reflections
correspond to perturbations of the vacua through colinear configurations
in the $W$ plane.  The affine models correspond to the case where $B = S
+ S^t$ has a single zero eigenvector, $v$.  Then $B$ defines a reduced
matrix $\hat{B}$ on the orthogonal complement to ${\bf R}v,$ which solves
the Diophantine equations.  Noting that $H^t v = -v,$ so $\lambda_v = 1$ and we
see that all the
eigenvalues $\lambda$ of $H$ have $\vert \lambda \vert = 1.$

\newsec{Orbifolds of ${\bf CP}^1.$}

Discrete subgroups of $SO(3)$ act naturally on ${\bf CP}^1,$ which is
topologically a sphere.  The description is
simplest in homogeneous coordinates.  The matrix
$\left(\matrix{a&b \cr c&d}\right) \in SU(2)$
sends the point $(x,y)$ to $(ax + by, cx +
dy).$  By the projective identification, the center $Z_2$ of $SU(2)$ acts
trivially, so $G \subset SU(2)$ acts by the image
under the covering $SU(2) \rightarrow SO(3).$  The topological orbifolds
of these models were considered in \rZAZ.  For genus zero
three-point functions,
from which the operator ring is derived,
we can use the sphere itself to represent a branched cover of the
(worldsheet) sphere, with the action of the group given by the
fundamental $SU(2)$ representation, as for the orbifold above.  Then
${\cal M}_k,$ the
holomorphic maps of degree $k,$ is represented by pairs of degree $k$
homogeneous polynomials:
\eqn\lmap{\Phi:  (X,Y) \mapsto \left({\sum_{l = 0}^{k}
\phi_{0l} X^{k-l}Y^{l},\sum_{l = 0}^{k} \phi_{1l} X^{k-l}Y^{l}}\right).}
So $\Phi$ is represented by a $2\times (k+1)$ matrix (defined up to
overall
multiplication by a scalar), acting on $(X^k,X^{k-1}Y,...,Y^k),$ and
the equivariant maps obey
\eqn\eqvar{\Phi \cdot \rho_k(g) = \lambda \rho_1(g) \cdot \Phi,}
where $\rho_1$ is the fundamental $SU(2)$ representation, and $\rho_k$ is
the $(k+1)-$dimensional representation on degree $k$ homogeneous
polynomials induced by $\rho_1$ ($\rho_k = (\otimes^k \rho_1)_{symm}$).
$\lambda$ is an arbitrary, possibly $g-$dependent factor.
Finding such maps amounts to finding intertwining maps of projective
representations \rM.\footnote{$^\dagger$}{Note: $\Phi$ must not be identically
zero.
Where the polynomials in \lmap\ have $r$ common roots, the roots are
divided out to get a lower degree map.  These maps, technically, are
in ${\cal M}_{k-r},$ not  ${\cal M}_{k},$ and make up the
compactification divisor.  The compactified
$\overline{{\cal M}}_{k}$ is then ${\bf CP}^{2(k+1)-1}.$}

Using the formula \twistcohom, we determine the ring of observables for
the $D_N$ to be
as follows.  If $N = 2k,$ there are two operators in the identity sector:
$1$ and $X,$ which descend from the original sigma model on ${\bf CP}^1.$
Associated to each conjugacy class $\theta_j$ of rotations
of the $N-$gon by $\pm 2\pi {j\over N},$ $j = 1...k-1,$
are two operators: call
them $\phi_j, \phi_{N-j}.$  The rotation by
$\pi$ is a central element of the group and has associated with it a
single operator, $\phi_k.$  In addition, there are two conjugacy classes
of flips (at a vertex or midpoint diagonal of the polygon),
which have one operator apiece,
$\rho$ and $\tau.$  The two fixed points under
$\rho = \left({\matrix{0&1\cr 1&0}}\right),$ for example,
are $(1,1)$ and $(1,-1)$ and are related by the $\pi$
rotation $\left({\matrix{i&0\cr 0&-i}}\right)$, which centralizes $\rho.$
The ring was shown in \rZAZ\ to be generated by
$\rho$ and $\phi \equiv \phi_1,$ and is given by
\eqna\genring{$$\eqalignno{\rho\phi^2 &= 4\rho&\genring a\cr
\rho^2 &= 1 + {1\over2}W_{2k}(\phi) +\sum_{l=1}^{k-1}W_{2l}(\phi)&\genring b\cr
\phi W_{2k}(\phi) &= 2W_{2k-1}(\phi),&\genring c\cr}$$}
where the functions $W_n(\phi)$ are the Chebyshev
polynomials, defined here to be
\eqn\cheb{W_n(X = 2\hbox{cos}(z)) = 2\hbox{cos}(nz).}
In terms of $\phi,$ the operators $\phi_j$ are
\eqn\phij{\phi_j = W_j(\phi).}
We also have $1 \equiv \half \phi_0 = \half W_0, \chi = \half W_N,$
and $\tau = \half \rho \phi$ (here $\chi$ = $X$ up to normalization by
the one-instanton action, i.e. the area of the target sphere:
$\chi = \beta^{-\half}X, \beta = \e{-A}).$

In the odd case, $N = 2k+1,$ there are again $N + 3$ operators.  The two
untwisted operators remain, as before; in each of the $k$ rotation
classes, there are two operators, $\phi_k, \phi_{N-k}$ (there is no
central element); and the lone flip conjugacy class has two operators,
$\rho$ and $\tau.$  The ring is then given by
\eqna\oddring{$$\eqalignno{\rho\phi^2 &= 4\rho&\oddring a\cr
\rho^2 &= {1\over2}W_{2k+1}(\phi) +\sum_{l=1}^{k}W_{2l-1}(\phi)&\oddring b\cr
\phi W_{2k+1}(\phi) &= 2W_{2k}(\phi),&\oddring c\cr}$$}
where the same expressions for the operators (as we've defined them) in
terms of $\phi$ and $\rho$ still hold.

In fact, both rings can be given
by the same equations.  Using the recursion relation obeyed by both
the $\phi_j$ and $W_j:$
$$xW_j(x) = W_{j+1}(x) + W_{j-1}(x) \qquad j \geq 1,$$
we can write the last equation as $W_{N+1}(\phi) = W_{N-1}(\phi).$
Writing $\phi = 2\hbox{cos}(z)$ this reads $2\hbox{cos}[(N+1)z]
= 2\hbox{cos}[(N-1)z],$ we can solve this equation as if it were
numerical, and get $z = {j \pi \over N},$ so the $x$ solutions are
$x_j = 2\hbox{cos}({j\pi\over N}).$
Most of
these roots ($j \neq 2,-2$), and $x = 0,$
are roots of the right hand side of \genring{b}\
{\sl and} \oddring{b}, from which we get
$$RHS \genring{b}\ \propto \phi \prod_{j = 1}^{N-1}(\phi - 2\hbox{cos}
(j\pi/N)) \propto {\phi \over \phi^2 - 4}(W_{N+1}(\phi) - W_{N-1}(\phi))$$
up to an overall constant, which can be determined by L'H\^opital's rule.
The result is the same in the odd and even case.  We use the last form of
the above equation and the recursion relation to write, for $N$ odd
{\sl or} even,
\eqna\orbring{$$\eqalignno{\rho\phi^2 &= 4\rho&\orbring a\cr
\rho^2 &= \half {W_{N+2}(\phi) - W_{N-2}(\phi) \over \phi^2 - 4}&\orbring b\cr
W_{N+1}(\phi) &= W_{N-1}(\phi).&\orbring c\cr}$$}
We stress that the right hand side of \orbring{b}\ is a polynomial.

In order to compute the matrix of soliton numbers associated to this
theory, it is convenient to work in a particular basis for the chiral
ring, the {\sl canonical} basis \rCVISING.  Such a basis
exists for finite $\beta.$  In this basis
the operator algebra reads
\eqn\canon{A_i\cdot A_j \propto \delta_{ij}A_j,}
where the constant of proportionality is determined by requiring that the
topological metric obeys
\eqn\canoneta{\eta_{ij} = \delta_{ij}.}
That $\eta$ is diagonal is a simple consequence of \canon.  To find this
basis, we use a trick associated to the chiral ring of a Landau Ginzburg
model (off criticality).
In particular, the point basis is obtained by writing the derivative of
the superpotential
$W^\prime(x) = \prod_{i}{(x - r_i)}$ and defining $A_j =
\prod_{i\neq j}(x - r_i)/W^\prime(r_j).$  The canonical basis differs only in
normalization.  In the above, the $r_j$ are complex numbers which satisfy
the ring relations.  In the Landau-Ginzburg case, they represent vacuum
expectation values of $W.$  We can perform this trick for our rings as
well, though the physical interpretation of the $r_j$ seems to be lost.

The numerical solutions to \orbring{}\ are as follows.  The first equation
implies $\rho = 0$ or $\phi^2 = 4.$  If $\phi^2 = 4,$ so $\phi =
2\epsilon,$ where $\epsilon = \pm 1,$ then since $2\epsilon =
2\hbox{cos}({\epsilon - 1\over 2}\pi),$ the last
equation reads $2\hbox{cos}({\epsilon-1\over 2}(N+1)\pi) =
2\hbox{cos}({\epsilon-1\over 2}(N-1)\pi))$ and is satisfied.  Equation
\orbring{b}\ can be
evaluated by L'H\^opital's rule to be
$\rho^2 = (-1)^{N({\epsilon - 1 \over 2})}N.$  The point basis, which
satisfies $A_i\cdot A_j = \delta_{ij}A_j,$ assigns to each solution a ring
element.  Let $(\rho_a,\phi_a)$ represent the $a^{th}$ solution
obtained above $(a = 1...N+3).$
The corresponding point basis element is then
\eqn\al{A_a =
{\prod_{\matrix{\sub i \sub , \sub j
\sub = \sub 1 \cr \sub \rho_i \sub\neq \sub \rho_a
\cr \sub \phi_j \sub\neq \sub \phi_a}}^{N+3}
{(\rho - \rho_i)(\phi - \phi_j)\over
(\rho_a - \rho_i)(\phi_a - \phi_j)}}.}
This expression can simplify greatly.  For example, if there is an
overall factor of $\rho,$ we can make the replacement $\phi^2 \rightarrow
4,$ by virtue of \genring{}.  It will be convenient for us to focus on the
$\phi$ pieces of the basis elements.  In fact, for all the points with
$\rho = 0,$ the ring elements only contain factors of $\phi.$  For the
other four points, $\rho^2 = \pm N,$ only the ``angular,'' i.e.
$\phi-$dependent, pieces are important for soliton number computations,
as we shall see below.  We thus define an ``effective'' basis
$\tilde{A}_a, a = 0...N,$ labeled by the $N+1$ values of $\phi: \phi_a =
2cos({a\pi \over N}).$  They have the simpler expression
\eqn\effal{\tilde{A}_a = \epsilon_a
{\prod_{\matrix{\sub i \sub = \sub 0 \cr
\sub i \sub\neq a}}^{N}{(\phi - \phi_i)\over (\phi_a - \phi_i)},}}
where $\epsilon_a = \half$ if $a = 0$ or $a=N,$ $\epsilon_a = 1,$ otherwise.

\newsec{Dynkin Diagrams and Dihedral Orbifolds}

In this section we will solve the $tt^*$ equations and compute the
soliton numbers between the vacua.  The resulting generalized Dynkin
diagram corresponds to the affine Lie group associated to the dihedral
group.

We recall a connection between discrete subgroups of $SU(2)$ and affine
Lie groups.  Each subgroup of $SU(2)$ has associated with it a
two-dimensional
fundamental representation, $R$.  The tensor product of any irreducible
representation, $V_i,$ with $R$ decomposes as
$$V_i \otimes R \cong \bigoplus_{j} A_{ij}V_j.$$
A theorem due to McKay \rSLOD\ states that the matrix $A$ is the adjacency
matrix of a Dynkin diagram for an
an affine Lie algebra.  For example, consider the fundamental (though
reducible) representation of $Z_N;$ the generator $g$ acts by
$g = \left({\matrix{\gamma &0 \cr 0 &\gamma^{-1}}}\right),$ where
$\gamma = \e{2\pi i/N}.$  The irreducible representations, ${\bf k}$
of $Z_N$ are just
$\gamma^k,$ and it is clear that ${\bf k} \otimes R \cong {\bf (k-1)}
\oplus {\bf (k+1)}.$  Now for a discrete subgroup of $SO(3),$ we use the
fundamental representation of the double cover of the subgroup.  So for
$Z_N$
we use $Z_{2N},$ and the same decomposition rule applies, with $k$
now ranging from zero to $2k-1.$  Thus we have the correspondence shown
in \fan.  The correspondence between the discrete dihedral groups
of $SO(3)$ and extended Dynkin diagrams of the $D-$series is
$$D_N \leftrightarrow \hat{D}_{N+2};$$
the $D_5 \leftrightarrow \hat{D}_7$ case is shown in \fdi.

Now we wish to compute the soliton matrix for the dihedral orbifolds.
By the discussion in section three, we know that we can compute a Dynkin
diagram from this object.
We will
compute it in the canonical basis, described in section four.
However, another basis is more convenient for solving the $tt^{*}$
equations, which we do below.

Implementing the symmetries of the orbifold simplifies the computation of the
ground state metric.  Namely, the product of all of the twists of
the operators must be the identity
(considered as incoming states, in genus zero).  For
nonabelian orbifolds, the product of two conjugacy classes contains a sum
of other conjugacy classes:  the decomposition of the classes is called
the group ring.  In order to have a nonzero correlation
function, the products of the classes in the group ring must contain the
identity.  For two point functions, this says that the metric is block
diagonal for conjugacy classes (the dual of $\{ h \}$ is $\{ h^{-1}\},$
which is $\{ h \},$ when viewed as an outgoing state).  Now no conjugacy
class contains more than two operators.  In addition to this constraint,
we have the reality constraint, i.e. $(CPT)^2 = 1.$  This can be used to
show that the metric is indeed diagonal, and in each sector has two real
positive components, $a,b,$ satisfying $ab = 1.$  Thus there is one real
parameter:
\eqn\tt{g_{\mu \overline{\nu}} = \left(\matrix{a &0 \cr 0
&a^{-1}}\right).}
We study behavior of $g$ as a function of the instanton action
$\vert \beta \vert = \e{-A}.$  Through a proper change of variables, all
of the equations are equivalent to the sinh-Gordon equation.
The $tt^*$ equation of interest, derivable from the flatness of \conns, is
$$\overline{\del}_{\overline{i}}(g\del_j g^{-1}) =
[C_j,\overline{C}_{\overline{i}}].$$
We are interested in the variations with respect to scale: $i = j
= \beta.$

The operator corresponding to $\beta$ is $-X/\beta.$  Since $X^2 \sim 1$
(and since its product with other fields always contains a single field)
we see that $C_\beta$ decomposes into $2\times2$ blocks;
acting on operators of fermion number ${k\over N}$ and
${N-k\over N}$ it has
the block form
$$C_\beta = - {1 \over \beta}\left(\matrix{0 &\beta^{k\over N} \cr
\beta^{N-k\over N} &0}\right)$$
(recall that $\beta$ can be assigned a chiral fermion number of two; then
this number is conserved in the ring products).  Note that for $N =
2k,$ the
operator $\phi_k$ obeys $X\phi_k = \beta^\half \phi_k,$ meaning that
$C_\beta$ is a $1\times1$ matrix in this block, giving that
$\langle \overline{\phi_k}\vert \phi_k \rangle$ is essentially constant
- it is a pure power of $\beta,$ due to our ``dimensionless'' definition
of fields.\footnote{$^\dagger$}{We have removed the appearance of $\beta$ in
the ring by defining ``dimensionless'' ring elements.  For example, in the
$k^{th}$ conjugacy class we have $\phi_k = \beta^{-k/2N}(\theta^{k}_A +
\theta^{N-k}_B),$ where $A$ and $B$ label the fixed points (likewise for
$\phi_{N-k}).$  We choose our sub-basis here so that $\eta =
\left(\matrix{0&1\cr1&0}\right).$  This was used in deriving \tt.}

Defining
\eqn\change{\matrix{x = 4|\beta|^{1\over2}, & \hbox{} & u(x) =
2\hbox{log }\left(a|\beta|^{{N-2k\over 2N}}\right)},} we get
\eqn\pthree{u^{\prime \prime} + {1\over x}u^{\prime} = 4\hbox{sinh }u.}
We must require finiteness of $g_{i\overline{j}}$ in the conformal $(\beta
\rightarrow 0)$ limit.  This tells us that as $x\rightarrow 0,$ $u$ behaves as
\eqn\rlim{u \rightarrow r\hbox{log}x + s, \qquad r = 2\left({N-2k\over
N}\right).}
The $x \rightarrow \infty$ behavior gives us the the matrix of soliton
numbers.  Namely, the metric should obey
\eqn\asymp{g_{i\overline{j}} \sim \delta_{ij} - {i\over
\pi}\mu_{ij}K_0(m_{ij}\beta).}
The $x \rightarrow \infty$ asymptotic behavior is known \rPIII.  The
solution to \pthree\ obeying \rlim\ contributes
\eqn\mupart{2i \hbox{sin}\left({\pi{N-2k \over 2N}}\right)\nu_k, \qquad \nu_k =
\cases{1, &$k = 0$\cr 2,&$k = 1...(N-1)$\cr 4,&$k = N$\cr}}
to $\mu$ in this expansion.  The factor $\nu_k$ is due to our choice of basis
(see footnote) and the fact that $X = \half \beta^\half W_{N}(\phi).$
Note that the operators corresponding to
flips all appear with ${k\over N} = \half$ (even when $N$ is odd), meaning
they do not contribute to the soliton numbers!  This was our
justification for isolating our analysis on the $\phi$ parts of the
canonical basis.

Before calculating the soliton
numbers, though, we wish to choose a standard basis -
the canonical basis, discussed in section four.  To do this, we express
the expansions of $A_l$ in terms of the above basis, which solved the
$tt^*$ equations.  We have that $\phi_k = \tilde{W}_k(\phi).$  So we
express products of the form \al\ as
\eqn\chebphin{\tilde{A}_l = \sum_{i = 0}^N{(c_l)_j \phi^j}
= \sum_{j = 0}^N (a_l)_j W_j(\phi).}
The following identity facilitates this change of basis.  From the
recursion relations, one can derive
\eqn\cob{x^n = \sum_{j = 1}^{\left[{n\over 2}\right]}\left({\matrix{n
\cr j}}\right)\tilde{W}_{n-2j}(x)}
(half of Pascal's triangle, in a way), where we have defined
$$\tilde{W}_n(x) = \cases{1, &$n=0,$\cr W_n(x),&$n>0.$\cr}$$
We still need to make one correction
to the normalization, so that $\eta_{ij} =
\delta_{ij}.$  This is simple since $X$ is the only operator with nonzero
topological correlation, and $\phi^n = \tilde{W}_n(\phi) + ...,$ where
$...$ represent lower degree polynomials (recall $2X \sim \tilde{W}_n).$
The result is that the canonical basis elements (just the $\phi$ parts),
which we denote $\hat{A}_l$ are
\eqn\canbas{\hat{A}_l = N_l\tilde{A}_l =
\pm\Big[{{1\over 2\epsilon_a}\prod_{\matrix{\sub i \sub = \sub 0 \cr
\sub i \sub\neq l}}^{N}(\phi_a - \phi_i)}\Big]^{\half}.}
Writing
\eqn\expand{\hat{A}_l = \sum_{i = 0}^{N}(\hat{a}_l)_i\tilde{W}_i(\phi),}
and using the diagonal property of the metric and
equations \asymp\ and \mupart,
we arrive at the expression for the soliton adjacency matrix
\eqn\solmat{\hat{A}_{rs} = \sum_{j = 0}^{N}(\hat{a}_r)_j(\hat{a}_s)_j
2i \nu_j \hbox{cos}\left({\pi{j \over N}}\right).}

The result is that the matrix $A,$ obtained from $\hat{A}$ by restoring the two
ring elements or $\hat{A}_0$ and $\hat{A}_N,$ is the adjacency matrix of the
extended
Dynkin diagram of the corresponding affine Lie group, up to choices of signs
for the $N_l.$  To be precise, the matrix $A,$ corresponding to the monodromy
in \monod\ is the upper triangular part of the matrix we obtained above.
We have checked our results explicitly for the first several values of
$N,$ and we obtain the expected form of $A,$ with integer (ones and
zeros) coefficients.  For higher $N,$ we have evaluated the expressions
numerically, and have obtained ones and zeros, though we know of no
mathematical proof that this must be so.  In the appendix, we check the
$N=5$ case explicitly.

Another check we can perform is obtaining $H$ from $A,$ i.e. $H =
(1-A)(1-A)^{-t},$ and computing its characteristic polynomial.  We know
the Ramond charges of the chiral ring, and these should be the phases of
the eigenvalues.  Indeed this is the case.

It is natural to guess that orbifolds by the exceptional discrete groups
are described by the exceptional affine Lie groups. Indeed, a simple
check of the characteristic polynomials of the matrices $H$ yields the
correct Ramond charges, though the full quantum ring and $tt^*$ equations have
not been computed.

\bigbreak\bigskip\bigskip\centerline{{\bf Acknowledgements}}\nobreak
I would like to thank C. Vafa for his valuable suggestions.  This work was
supported in part by Fannie and John Hertz Foundation and by NSF contract
PHY-92-8167.

\app{A}{}
In this appendix, we show a detailed computation of
some of the matrix elements of $A$ for the $D_5$ orbifold
of ${\bf CP}^1.$

By taking the real part of $\sum_{j=1}^{5}\e{2\pi i j/5} = 0,$ one easily
calculates
$$\eqalign{&x \equiv \hbox{cos}(\pi/5) = {1+\sqrt{5}\over 4}\cr
&y \equiv \hbox{cos}(2\pi/5) = {-1+\sqrt{5}\over 4}}.$$
Note $xy = 1/4;$ $x^2 + y^2 = 3/4;$ $x^2 - y^2 = \sqrt{5}/4;$ $2x^2 = 1+y;$
$2y^2 = 1-x.$
Then, using \al, \canbas, and \cob\ we get
\eqn\azero{\eqalign{\hat{A}_0 &= N_0\epsilon_0\tilde{A}_0 \cr
&= N_0\half\prod_{l=1}^{5}(\phi - 2\hbox{cos}(l\pi/5))/(2 -
2\hbox{cos}(l\pi/5))\cr
&= N_0\half(\phi - 2x)(\phi - 2y)(\phi + 2y)(\phi + 2x)(\phi + 2)/20 \cr
&= \pm{\sqrt{20}\over 40}(\phi+2)(\phi^2 - 4x^2)(\phi^2-4y^2) \cr
&= \pm{\sqrt{5}\over 20}[\phi^5 + 2\phi -3\phi^3 - 6\phi^2 + \phi + 2] \cr
&= \pm{\sqrt{5}\over 20}[W_5(\phi) + 2W_4(\phi) + 2W_3(\phi) + 2W_2(\phi)
+ 2W_1(\phi) + 2].}}
Similarly, making use of the identities above, one finds
\eqn\aone{\hat{A}_1 = \pm{i\sqrt{5}\over 10}[W_5(\phi) + 2xW_4(\phi) +
2yW_3(\phi) - 2yW_2(\phi) -2xW_1(\phi) - 2].}
Plugging into \solmat yields
\eqn\amat{\eqalign{A_{01} &= \pm \left({\sqrt{5}\over 20}\right)
\left({\pm{i\sqrt{5}\over 10}}\right)(2i)
[1\cdot 4\hbox{cos}(5\pi/5) + 4x\cdot 2\hbox{cos}(4\pi/5)\cr
& \qquad \qquad \qquad \qquad \qquad \qquad+ 4y\cdot 2\hbox{cos}(3\pi/5)
-4y\cdot 2\hbox{cos}(2\pi/5) \cr
& \qquad \qquad \qquad \qquad \qquad \qquad -4x\cdot 2\hbox{cos}(\pi/5)
-4\hbox{cos}(0)]\cr
&= \pm\left({1\over 5}\right)
[\hbox{cos}(\pi) + 2x\hbox{cos}(4\pi/5)
+ 2y\hbox{cos}(3\pi/5) \cr
& \qquad \qquad -2y\hbox{cos}(2\pi/5) - 2x\hbox{cos}(4\pi/5) -
\hbox{cos}(0)]\cr
&= \pm\left({1\over 5}\right)[-1 - 2x^2 - 2y^2 - 2y^2 - 2x^2 - 1]\cr
&= 1,}}
where we have used the freedom of signs to choose $+1.$

One finds $A_{12} = A_{23} = A_{34} = A_{45} = 1$ as well. Recalling that there
are really two ring elements corresponding to both $\hat{A}_0$ and $\hat{A}_5$
gives us the Dynkin diagram of $\hat{D}_8,$ shown in \fdi.  Specifically, to
solve the Diophantine equations, we take the upper triangular matrix $A_{i<j},$
remembering that $\hat{A}_0$ and $\hat{A}_N$ each represent two operators.
Specifically, for $D_5,$ we obtain
\eqn\fulla{A = \left(\matrix{
0&0&1&0&0&0&0&0\cr
0&0&1&0&0&0&0&0\cr
0&0&0&1&0&0&0&0\cr
0&0&0&0&1&0&0&0\cr
0&0&0&0&0&1&0&0\cr
0&0&0&0&0&0&1&1\cr
0&0&0&0&0&0&0&0\cr
0&0&0&0&0&0&0&0}\right).}
The monodromy matrix is
$$H = (1-A)(1-A)^{-t}$$
and its characteristic equation is
$$\eqalign{\hbox{det}(z - H) &= z^8 - z^7 - z^6 + z^5 +z^3 - z^2 -z -1\cr
&= (z-1)^2(z+1)^2(z^4-z^3+z^2-z+1)\cr
&= \Psi_1(z)^2\Psi_2(z)^2\Psi_{10}(z),}$$
where $\Psi_n(z)$ is the $n^{th}$ cyclotomic polynomial.  We can easily read
off the Ramond charges of the superconformal theory to be
$$\{ q_i \} = \left\{ { -\half, -{3\over 10}, -{1\over 10}, 0,0,{1\over
10},{3\over 10},\half }\right\},$$
which are the NS charges shifted by $-{\hat{c}\over 2} = -\half,$ as expected.

\listrefs
\listfigs

\bye